\documentstyle[12pt,epsf]{article}

\long\def\capfont{\@setsize\capfont{10pt}\xpt\@xpt}

\long\def\@makecaption#1#2{%
   \vskip 10\p@
   \setbox\@tempboxa\hbox{\capfont#1: #2}%
   \ifdim \wd\@tempboxa >\hsize
   \begin{quote}
     \capfont  #1: #2\par
   \end{quote}
     \else
       \centerline{\hfil\box\@tempboxa\hfil}%
   \fi}

\begin{document}

\thispagestyle{empty}
\title{Absence of Parity-Flavor Breaking Phase in QCD With Two Flavors of
Wilson Fermions for $\beta \geq 5.0$.\thanks{Submitted to Physical Review D.}} 

\author{Khalil M.~Bitar \\
Supercomputer Computations Research Institute, \\
Florida State University\\
Tallahassee,  Florida 32306-4052\\
U.S.A.\\[3ex]
FSU-SCRI-96-11\\[4ex]
}

\date{\today}
\maketitle

\begin{abstract}
We present data testing the existence of a parity-flavor breaking
phase in simulations of QCD with two flavors of light Wilson fermions.
This is done by explicit simulations on lattice sizes of $6^4$, $8^4$
and $10^4$ for a variety of values of $\beta$ and $\kappa$ as well as
the coefficient, $h$, of an explicit breaking term included in the
action.  We find that at $\beta=6/g^2 $ equal to or greater than 5.0
extrapolation in the parameter $h$ as well as in the lattice volume show
no indication of a phase where parity and flavor are spontaneously
broken in the limit of zero $h$.
\end{abstract}


\newpage
\section{Introduction}
\label{introduction}
For many years now, Aoki~\cite{Aoki84a,Aoki84b,Aoki86,Aoki89,Aoki95}
and collaborators~\cite{Gocksch89,Gocksch90,Gocksch92} have been
advocating the existence of a parity-flavor breaking phase in QCD with
Wilson fermions as a means of explaining why the pion mass in this
model approaches small values as the Wilson parameter $\kappa$
approaches, for every value of inverse square coupling $\beta$, a
critical value $\kappa_c$. This in spite of no-go
theorems~\cite{Vafa84a,Vafa84b}
that forbid such a phase in the continuum limit.

 Indeed analytic arguments have been presented to support the
existence of such a phase at $\beta=0.0$. For finite and, in
particular, larger values of $\beta$ where current lattice simulations
are undertaken, such evidence is lacking~\cite{Gocksch92}.

 Although the picture advocated by Aoki may explain the smallness of
the pion masses as $\kappa$ approaches $\kappa_c$ it also explicitly
states that these pions are not the Goldstone modes of spontaneous
chiral symmetry breaking. This presents a problem in that it is not
then clear that any of the soft pion and other theorems associated
with this phenomenon will be respected on the lattice. In other words
this would not be the expected simulation of true QCD.  In fact the
large $N$ analytic analysis which indicates the existence of this phase
at $\beta=0.0$ also shows the non-vanishing of the $\pi - \pi$
scattering length which is contrary to the the expected spontaneous
chiral symmetry breaking of QCD. This is of course unimportant at
$\beta=0.0$ but is very important, if true, at the values of $\beta $
where current simulations are performed.

 The alternative picture where the explicit chiral symmetry breaking
Wilson term causes the (otherwise Goldstone) pions to acquire a small
mass proportional to the lattice spacing does not have such a problem.
Indeed, that all these extra effects would disappear as the lattice
spacing is made smaller with the approach to the continuum limit was
formally demonstrated some time
ago~\cite{Karsten81,Seiler82,Bichicchio85}.

Several models exhibit a parity-flavor breaking phase. In the
Nambu-Johna-Lassinio~\cite{Bitar94a,Bitar94b} model with Wilson
fermions this phase was numerically and, in the large $N$ approximation,
analytically \cite{Bitar94a,Gocksch94} confirmed for values of
$\beta$ up to a specific cutoff value. This phase disappears for
larger values of $\beta$. The Schwinger model with two flavors of
Wilson fermions exhibits this phase at strong coupling and also loses
it at weak coupling~\cite{Horvath95}.  Whereas the picture advocated
by Aoki does not anticipate such a quenching effect for the phase in
QCD, recent phenomenological arguments by Creutz~\cite{Creutz95} tend
to show a preference for quenching of this phase if it exists.

Thus it becomes necessary to explore this important feature by
explicit simulations of QCD with Wilson fermions on volumes larger
than those already studied~\cite{Gocksch92}.

\section{Signature of the Broken Phase}
\label{Signature of the Broken Phase}

\begin{figure}
\vspace{3.25in}
\includegraphics{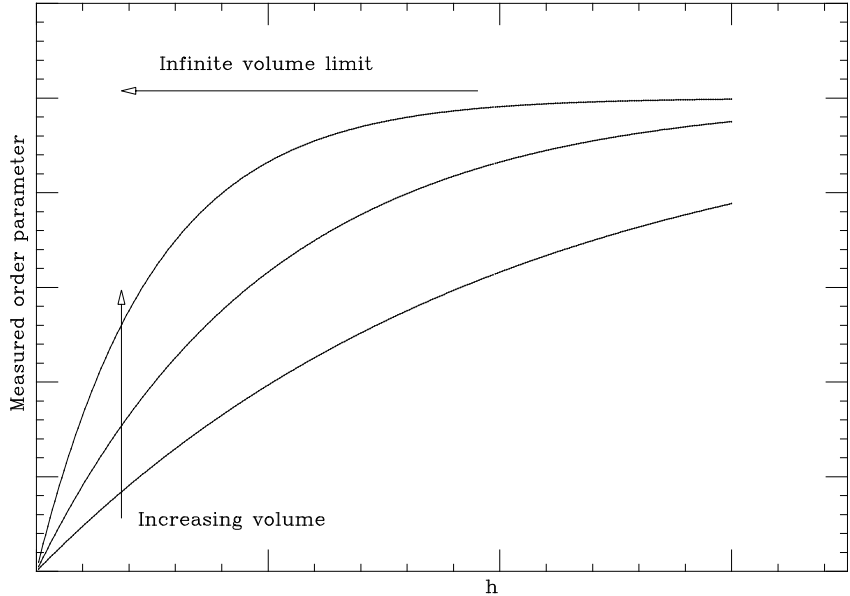}
\caption{Expected variation of computed order parameter with volume.}
\label{fig1}
\end{figure}

 Following arguments presented by Aoki and Aoki and
Gocksch~\cite{Aoki84a,Gocksch92}, it is necessary, in order to
investigate the presence of the parity-flavor-breaking phase in
simulations, to introduce first an explicit breaking term into the
action and then extrapolate the measured order parameter as this term
tends to zero.  Since the extrapolation is to be done, in principle,
after the `infinite volume' limit is taken, such simulations must be
done for larger volumes and any order parameter extrapolation be
studied as a function of this increasing volume. In such a situation a
typical behaviour of the data at finite volumes that one might expect
is shown in Fig.~\ref{fig1}. It is expected that the functional dependence of
the (measured) order parameter on $h$ be such that for any $finite$
volume this order parameter vanishes at $h =0.0$.

 The variation of this dependence with increasing volume is crucial to the
initial determination of the existence of a broken phase or its absence.
The existence of a broken phase in the infinite volume limit is signaled
by a flattening of this dependence for larger values of $h$ and a sharper
drop to zero as $h$ approaches zero. Thus it is clear that a significant
volume dependence of the order parameter at smaller values of $h$ is a
necessary indicative factor for this phase.  If, on the other hand, the
approach to zero is not varying significantly as the volume increases, the
infinite volume limit will not sustain a broken phase.\footnote{If one is
not careful one may arrive at wrong conclusions.  If simulations are done
only at larger values of $h$ one may use the slightly varying values of
the order parameter there to extrapolate these to $infinite$ $volume$. If
this step is then followed by a linear extrapolation to $h=0.0$ a non-zero
value for the order parameter at $h=0.0$ may be obtained.  At this stage
it is tempting to conclude that a broken phase exists in that limit. This
may be the wrong conclusion if this is not accompanied by a significant
increase in the value of the order parameter at smaller values of $h$.}

\section{Numerical Simulations}
\label{Numerical Simulations}

 We report here on simulations done with two flavors of Wilson
fermions at $\beta=5.0, 5.5$, and $8.0$ on volumes of $6^4, 8^4$, and
$10^4$ for a variety of values of $\kappa$ ranging from less than the
appropriate $\kappa_c$ to values greater than $\kappa_c$.

 The choice of these three values of $\beta$ was determined as follows.
The value $\beta=5.5$ represents current simulations on larger lattices
where spectrum and matrix element calculations are being done; that at
$\beta=5.0$ represents a lower value below which relevance to continuum
physics is not expected, and the last value at $\beta=8.0$ is to extend
the search to a much larger value of $\beta$ in case the parity-flavor
breaking phase were to be confirmed at the two smaller values.

 We introduce into the QCD action a term of the form $i h \bar{\psi}
\gamma_5 \tau_3 \psi$ where $\tau_3$ is a $2\times2$ matrix representing the
third element of the generators of flavor SU(2) algebra.  Upon integrating
the fermionic variables this is reflected in the simulation by the product
of two determinants: $Det M(h) * Det M(-h) $ where $M(h)$ is given by a
simple modification of the Wilson Matrix $M_w$ as:

   $$ M(h) = M_w + ih\gamma_5$$
As pointed out by Aoki, we also have here 
   $$ \gamma_5 M(-h) \gamma_5 = M^\dagger $$
and:
   $$ Det M(-h) = Det M^\dagger (h)\, . $$

Simulations were done for the parameter $h$ taking values ranging from 0.001 to
0.3. For the volume dependence we concentrate on the smaller values of $h$ and
in particular $ h= 0.001$ and $h= 0.005$ for all three volumes considered and
mostly for values of $\kappa$ greater than $\kappa_c$.

 The order parameter we compute is the expectation value of the operator $ i
\bar{\psi} \gamma_5 \tau_3 \psi$. With our notation this is given as

      $$ PF= Im Tr(\gamma_5 M^{-1}(h)) $$

\section{Results}
\label{Results}

   For the three values of $\beta$ considered,  simulations were performed,  as
mentioned above,  at
various values of $\kappa$ both below and above $\kappa_c$. We shall present
the data and results for each value of $\beta$ considered separately.

In all cases these simulations were also done at various values of the
external parameter $h$.  For each $\kappa$ the results of the
compuations on the three volumes
 $L^4$, $L=6, 8$, and $10$ were compared at the two values of
$h=0.001$, and $h=0.005$. The variation of the order parameter with
$1\over L$ is then used to obtain an `infinite volume' limit for all
values of $h$ used. The choice of $1\over L$ is indicated here by the
naive dimension of the order parameter. Following this, the order
parameter at these values of $h$ were fitted to:
$$ PF = A + B h^{1\over3}+ C h + D h^2 $$ 
 A separate fit to the pure quadratic polynomial 
$$ PF = A + C h + D h^2 $$ 
was also done.

 The initial aim in this case is to detect the possible existence of any
non-zero constant $A$ at $h=0.0$ as the limit of the order parameter at that
point. This is of particular interest for comparing results at values of
$\kappa$ above $\kappa_c$ with those below $\kappa_c$.  

It is useful to point out here that in the presence of a parity-flavor
breaking phase the order parameter is expected to vary with $h$ as:

$$ PF_\infty = A + B h^{1\over3} +\ldots\, ,$$
this being the behaviour of the root of the cubic equation determining the
position of the minimum of the quartic effective potential. In the absence of
such a phase the same behaviour follows with $A=0.0 $. As the quartic potential
becomes quadratic the leading behaviour becomes: 
$$ PF = C h + \ldots$$  

This should,  when compared to the data, be also a useful tool in determining
which situation one is in.

\subsection{$\beta=5.0$}
    
    The value of $\kappa_c$ at this value of $\beta$ is known to be about 
 $0.18$. We consequently performed simulations well below
that value at $\kappa=0.15 $ and well above it at $\kappa=0.1875 $
and intermediate values in between. We present in  table~\ref{table-one} the
results of these  simulations.     

\begin{table}
\caption{Parameters and measured order parameter $PF_L$ for the
 case of $\beta=5.0$ on lattices of volume $L^4$,  for $L=6$, $8$, and
$10$.}
\begin{center}
    \begin{tabular}{ccccc}
      \hline
       $\kappa$ & $h$ & $PF_6$ & $ PF_8$ & $PF_{10}$\\
      \hline
       0.1500 & 0.001 & 0.01969(31) & 0.01966(17) & 0.01966(11) \\
       0.1500 & 0.005 & 0.0983(15)  & 0.09834(86) & 0.09835(51) \\
       0.1500 & 0.050 & 0.968(15)   &             &             \\
       0.1500 & 0.100 & 1.863(26)   &             &             \\
       0.1500 & 0.300 & 4.310(48)   &             &             \\
      \hline
       0.1810 & 0.001 & 0.0259(20)  & 0.0277(22)  & 0.0273(2)   \\
       0.1810 & 0.005 & 0.1319(90)  & 0.1294(57)  & 0.1365(80)  \\
      \hline
       0.1820 & 0.001 & 0.0255(17)  & 0.02629(98) & 0.02559(8)  \\
       0.1820 & 0.005 & 0.1282(87)  & 0.1325(74)  & 0.1315(53)  \\
       0.1820 & 0.050 & 1.396(56)   & 1.389 (29)  &             \\
       0.1820 & 0.100 & 2.382(56)   & 2.380 (32)  &             \\
       0.1820 & 0.300 & 4.619(58)   &             &             \\
      \hline
       0.1850 & 0.001 & 0.0250(13)  & 0.0250(7)   & 0.0251(5)   \\
       0.1850 & 0.005 & 0.1225(52)  & 0.1273(71)  & 0.1223(18)  \\
       0.1850 & 0.050 & 1.287(65)   &             &             \\
      \hline
       0.1875 & 0.001 & 0.0239(8) & 0.0243(6) & 0.0240(2) \\
       0.1875 & 0.005 & 0.1193(42) &0.1206(23) & 0.1201(16) \\
      \hline
    \end{tabular}
    \label{table-one}
\end{center}
\end{table}

\begin{figure}
\vspace{3in}
\includegraphics{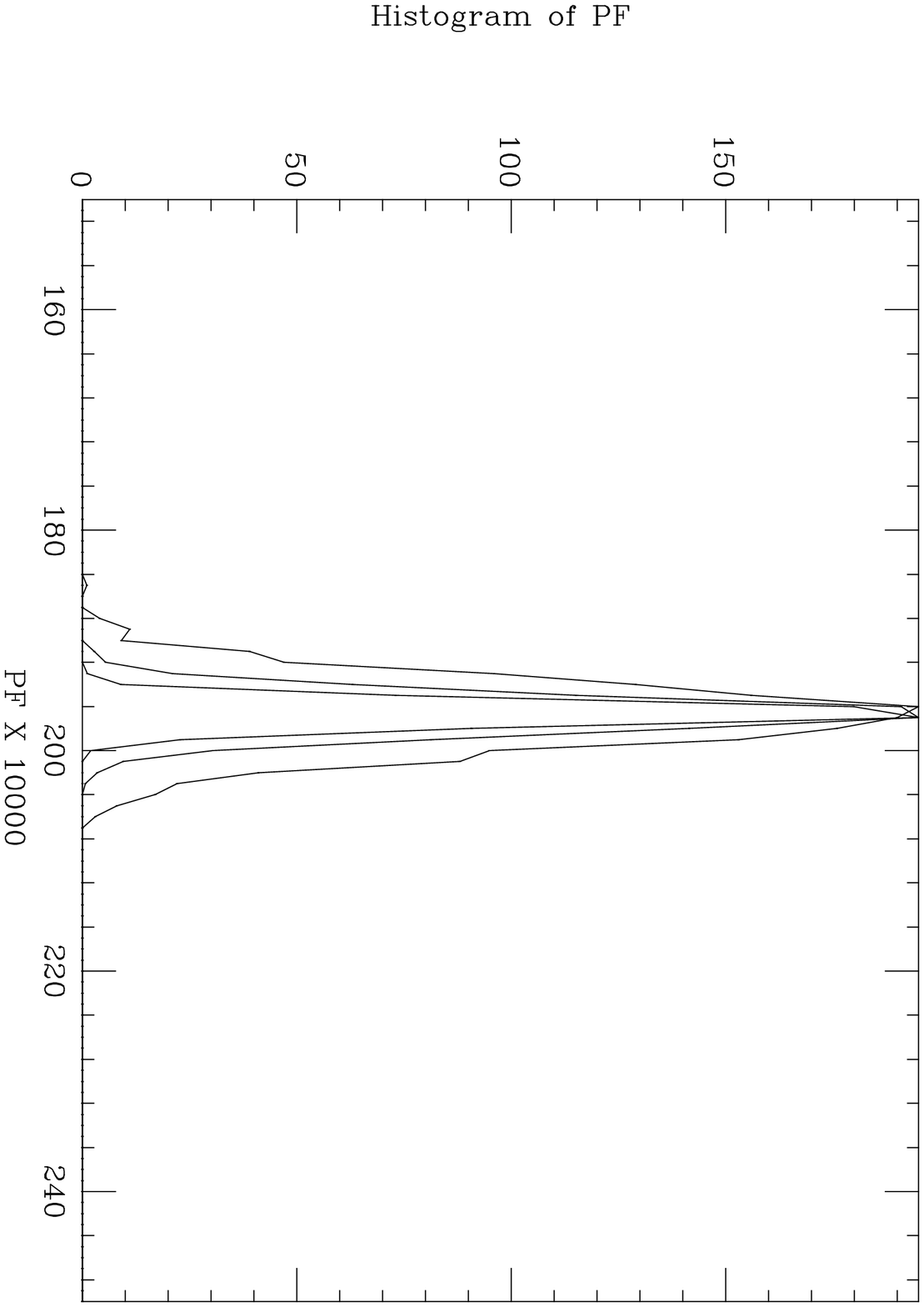}
\includegraphics{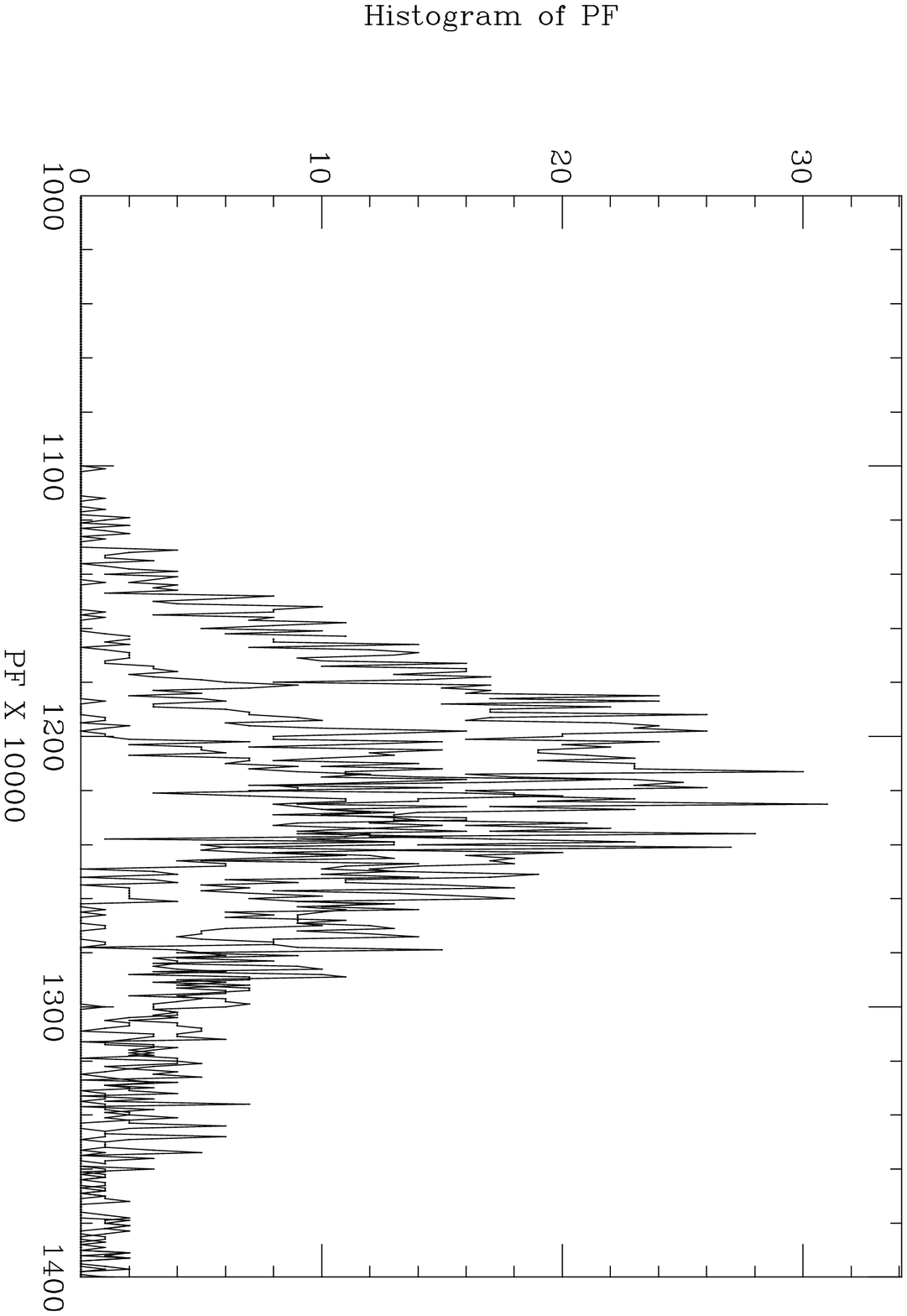}
\caption{(a) Histogram of computed PF at $\beta= 5.0$ $\kappa=0.15$
$h=0.001$ for all volumes considered; and 
(b) Histogram of computed PF at $\beta= 5.0$ $\kappa=0.185$ $h=0.005$ for
all volumes considered.}
\label{fig2}
\end{figure}

\begin{figure}
\vspace{5in}
\includegraphics{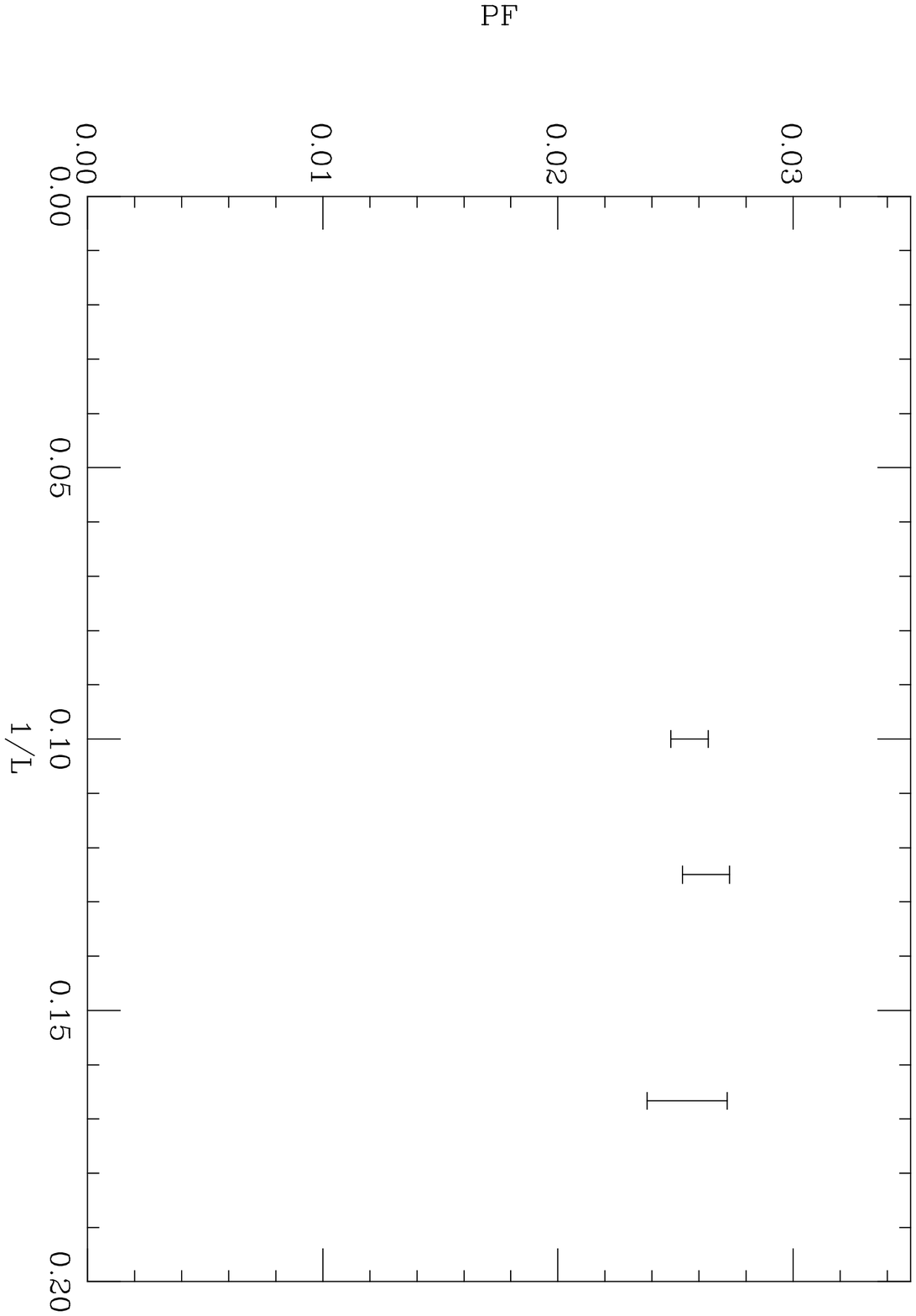}
\includegraphics{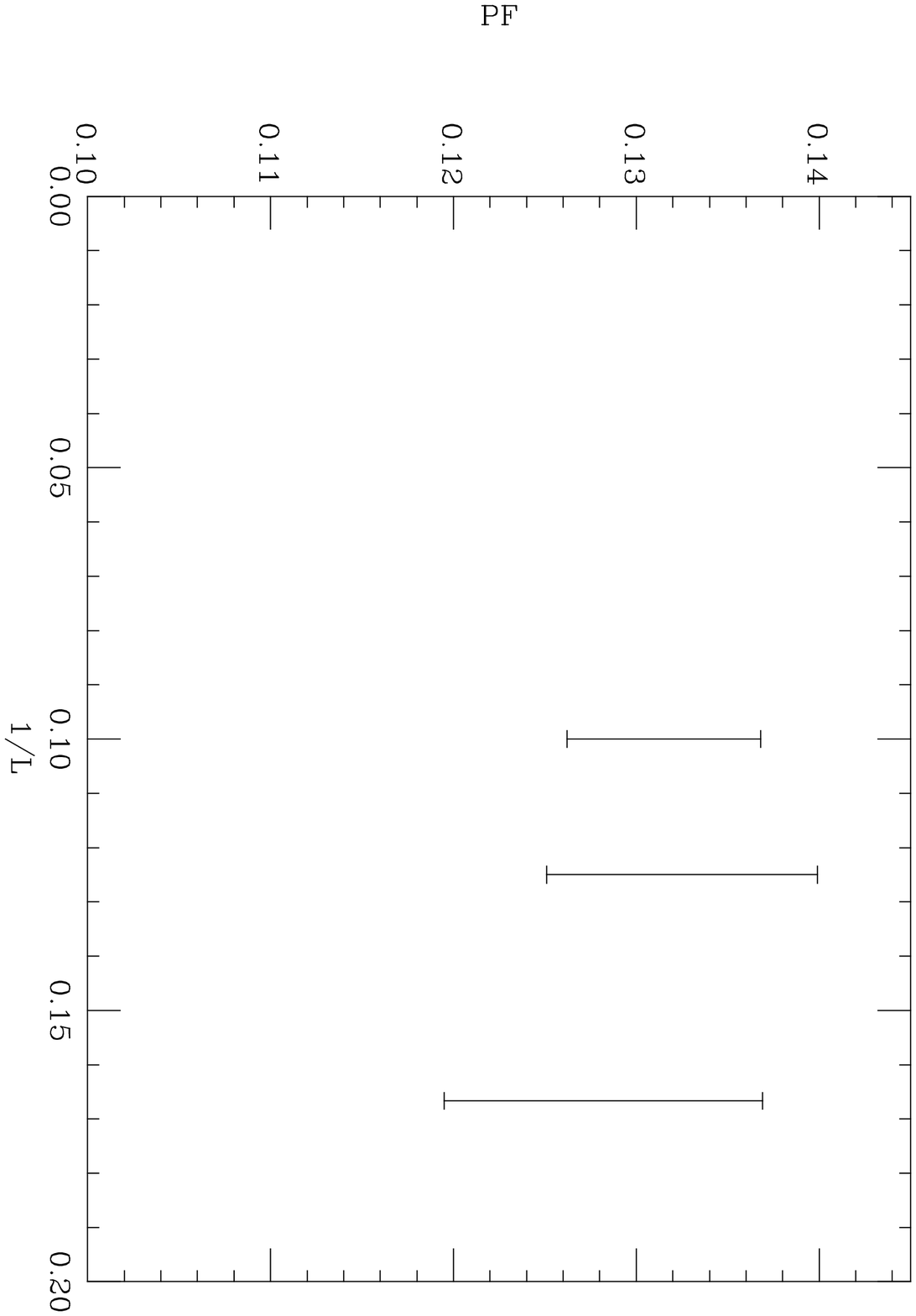}
\includegraphics{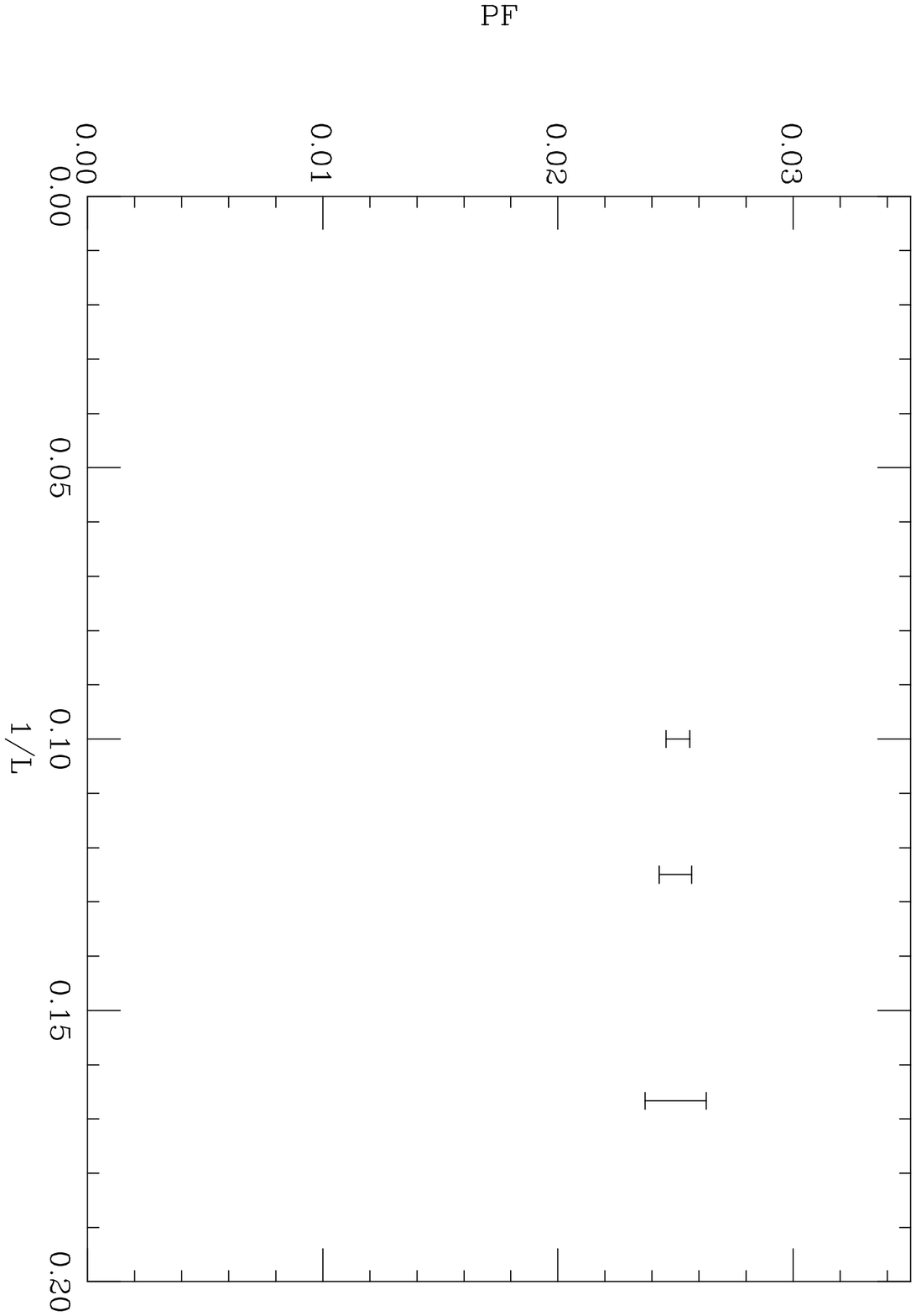}
\includegraphics{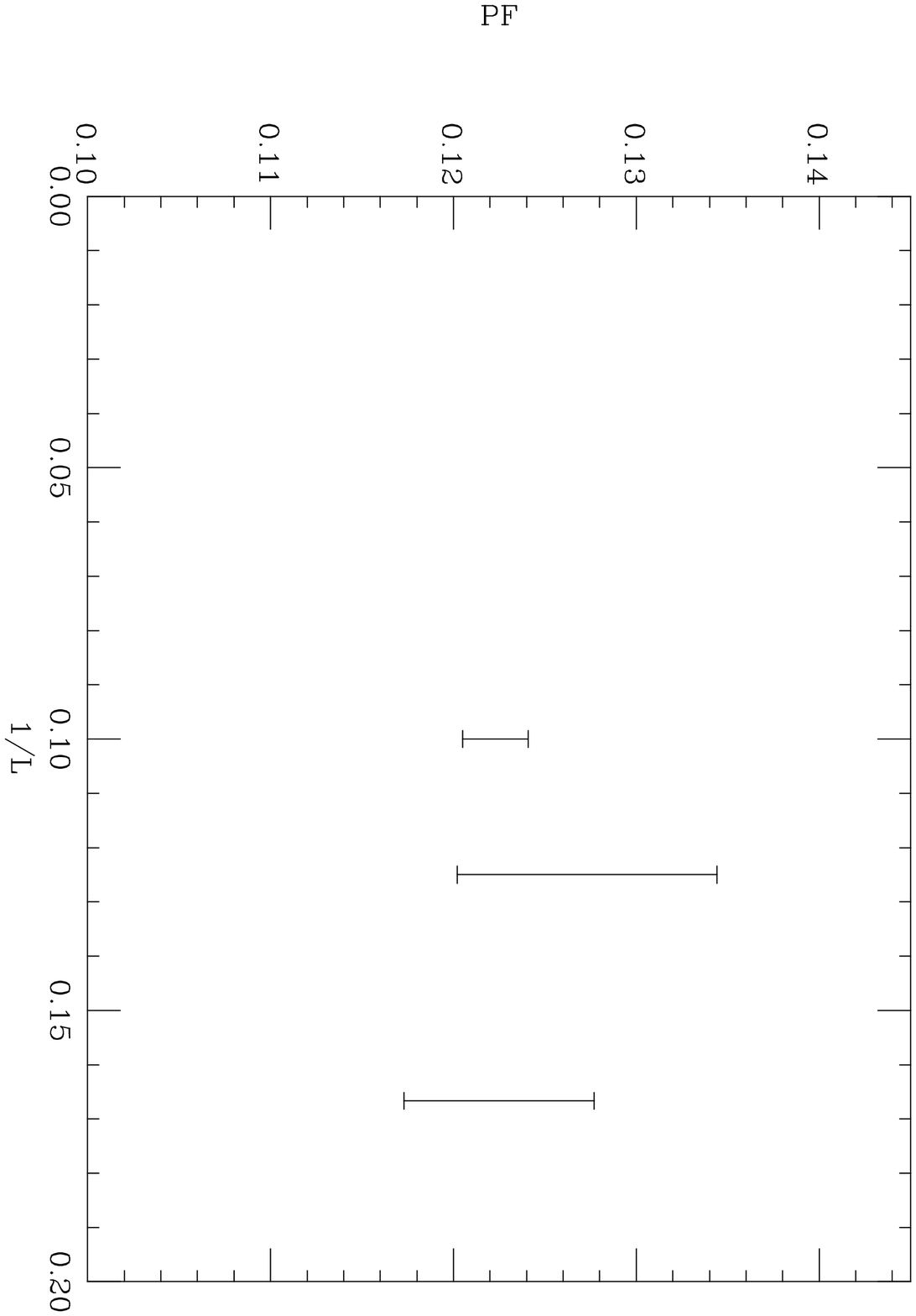}
\caption{(a) PF vs. $1\over L$ for $\beta=5.0$ $\kappa=0.182$ $ h=0.001$;
(b) PF vs. $1\over L$ for $\beta=5.0$ $\kappa=0.182$ $ h=0.005$;
(c) PF vs. $1\over L$ for $\beta=5.0$ $\kappa=0.185$ $ h=0.001$; and
(d) PF vs. $1\over L$ for $\beta=5.0$ $\kappa=0.185$ $ h=0.005$.}
\label{fig3}
\end{figure}

\begin{figure}
\vspace{3in}
\includegraphics{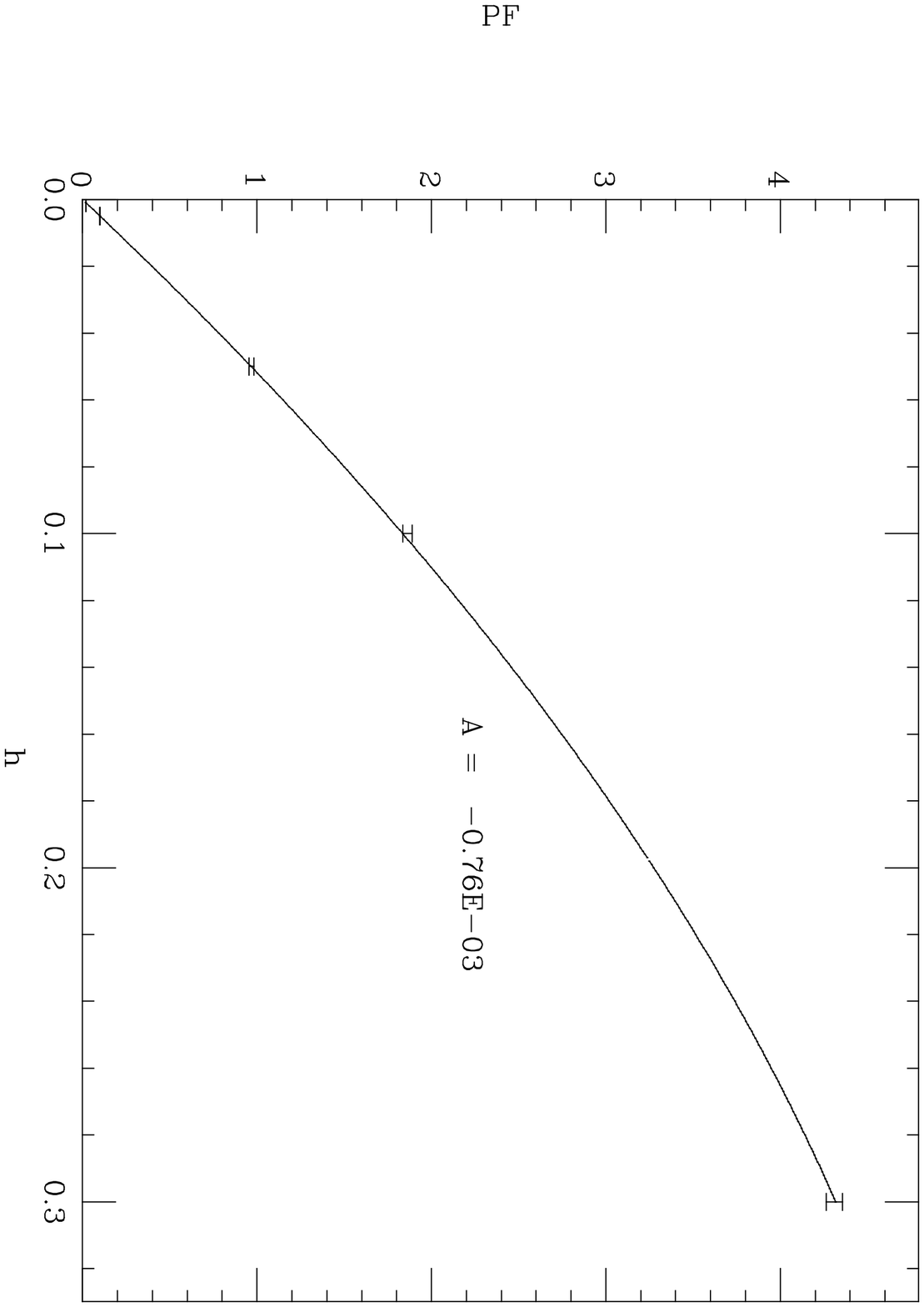}
\includegraphics{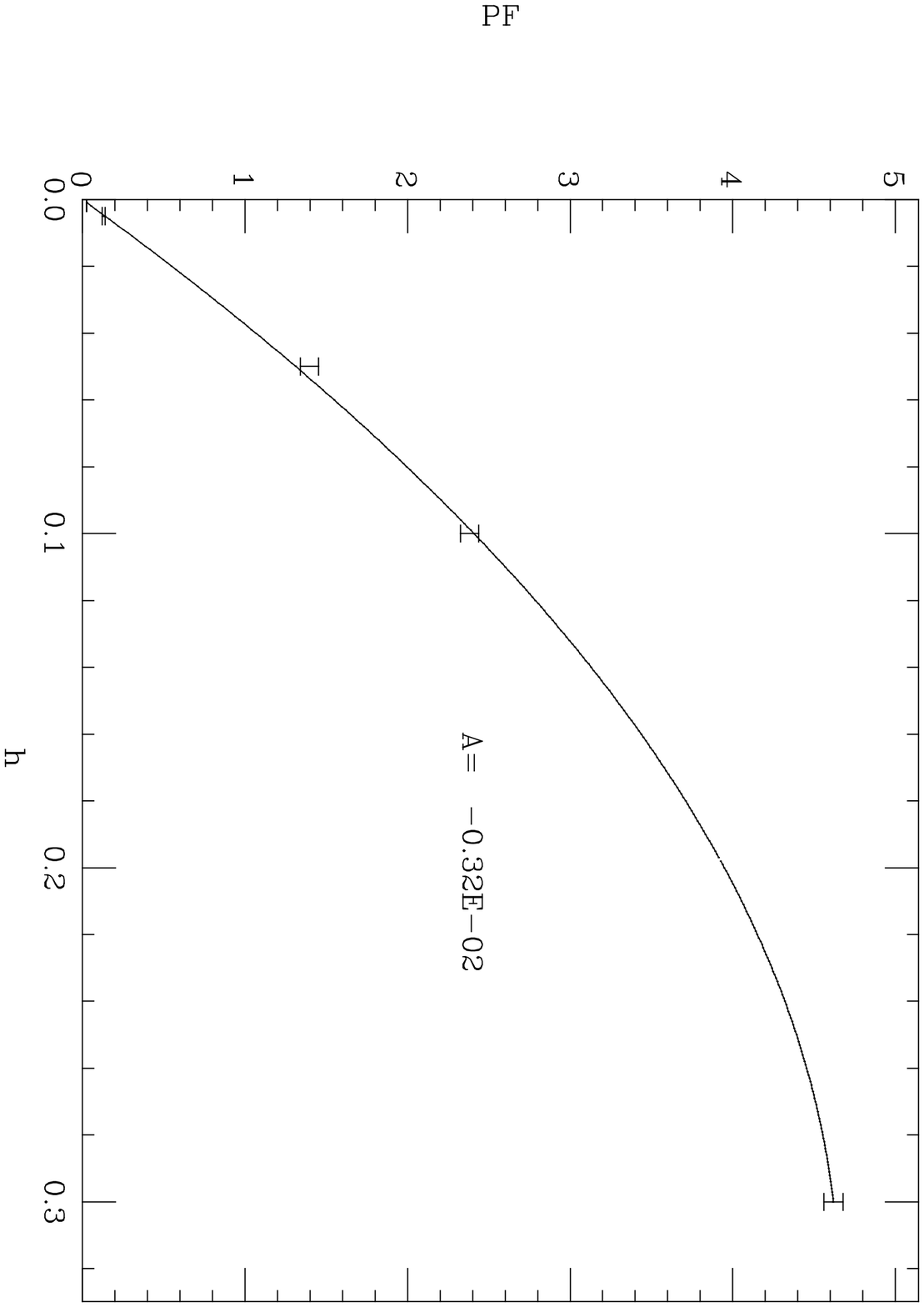}
\caption{(a) PF vs. $h$ for $\beta=5.0$ $\kappa=0.15$ and a quadratic
fit; and (b) PF vs. $h$ for $\beta=5.0$ $\kappa=0.182$ and a quadratic fit.}
\label{fig4}
\end{figure}

 The results in table~\ref{table-one} clearly show also that the
values computed for the order parameter at the larger values of the
volume are only incrementally different from those measured on the
small volume for all values of $\kappa$ indicated.
 For values of $\kappa$ less than $\kappa_c$, these results are
consistent within errors. This is best illustrated by the overlapping
histograms of these measurements at $\kappa=0.15$ given in
Fig.~\ref{fig2}a. For $\kappa = 0.185$ a similar histogram,
Fig.~\ref{fig2}b, indicates only an incremental increase of the peak
of the distribution with volume. This incremental change may be used
to obtain an `infinite volume limit' of these values assuming a linear
extrapolation in $1\over L$ where $L$ is the lattice linear dimension
as shown for example in Figs.~\ref{fig3}a, b, c, and d.  It is clear
here that the data is consistent with being essentially `constant'
with volume.  A quadratic fit in $h$ to this `infinite volume values'
is not significantly different from a fit to the data at volume $6^4$
where we obtain -- for example, Figs.~\ref{fig4}a and b -- a zero
constant for the extrapolated value of the order parameter at $h=0.0$
for $\kappa=0.15 $ below $\kappa_c$ and $\kappa=0.182$ slightly above
it.  For values of $\kappa $ both below and above $\kappa_c$ this clearly
implies the absence of any volume dependence of the order parameter
and, hence, in both cases and in particular the latter case, the
absence of a flavor-parity breaking phase in the system at $\beta
=5.0$. A fit with a leading $h^{1\over 3} $ is not as a good a description
of the data as it leads to a much higher $\chi-square$. Hence, we further
conclude that the effective potential of the system is
predominanty quadratic at small $h$.

\subsection{$\beta=5.5$}

   The value of $\kappa_c$ in this case is also known to be in the
neighborhood of $\kappa = 0.16$.  Table~\ref{table-two} details the
results of our compuations for values of $\kappa$ well below and above
this value. We concentrate in this discussion on the results obtained
at $\kappa=0.162$ and $0.165$ both above $\kappa_c$ and where the
postulated phase is expected to exist.

\begin{table}
\caption{Parameters and measured order parameter $PF_L$ for the
 case of $\beta=5.5$ on lattices of volume $ L^4$,  for $L=6, 8, 10
$.}
\begin{center}
    \begin{tabular}{ccccc}
      \hline
       $\kappa$ & $h$ & $PF_6$ & $ PF_8$ & $PF_{10}$\\
      \hline
       0.1300 & 0.001 & 0.01689(22) &             &             \\
       0.1300 & 0.005 & 0.0844(10)  &             &             \\
       0.1300 & 0.050 & 0.836(10)   &             &             \\
       0.1300 & 0.100 & 1.633(20)   &             &             \\
      \hline
       0.1350 & 0.001 & 0.01750(23) & 0.01750(14) &             \\
       0.1350 & 0.005 & 0.0874(12)  & 0.0875(7)   &             \\
      \hline
       0.1425 & 0.001 & 0.01860(29) & 0.01863(17) &             \\
       0.1425 & 0.005 & 0.0931(15)  &             &             \\
      \hline
       0.1500 & 0.001 & 0.02000(40) & 0.02009(24) &             \\
       0.1500 & 0.005 & 0.0998(20)  &             &             \\
      \hline
       0.1550 & 0.001 & 0.02079(51) & 0.02140(36) &             \\
       0.1550 & 0.005 & 0.1039(26)  &             &             \\
      \hline
       0.1610 & 0.005 & 0.1071(28)  & 0.1112(25)   & 0.1120(21)  \\
      \hline
       0.1620 & 0.001 & 0.02132(54) & 0.02200(42) & 0.02246(31) \\
       0.1620 & 0.005 & 0.1068(26)  & 0.1102(20)  & 0.1114(17)  \\
       0.1620 & 0.050 & 1.041(24)   & 1.052(14)   &             \\
       0.1620 & 0.100 & 1.954(36)   & 1.969(22)   &             \\
       \hline
       0.1650 & 0.001 & 0.02158(6)  & 0.02205(43) & 0.02299(29) \\
       0.1650 & 0.005 & 0.1086(34)  & 0.1101(20)  & 0.1110(15)  \\
       0.1650 & 0.050 & 1.047(23)   & 1.056(14)   &             \\
       0.1650 & 0.100 & 1.966(35)   & 1.984(21)   &             \\
       0.1650 & 0.300 & 4.354(50)   &             &             \\
      \hline
    \end{tabular}
    \label{table-two}
\end{center}
\end{table}

\begin{figure}
\vspace{3.25in}
\includegraphics{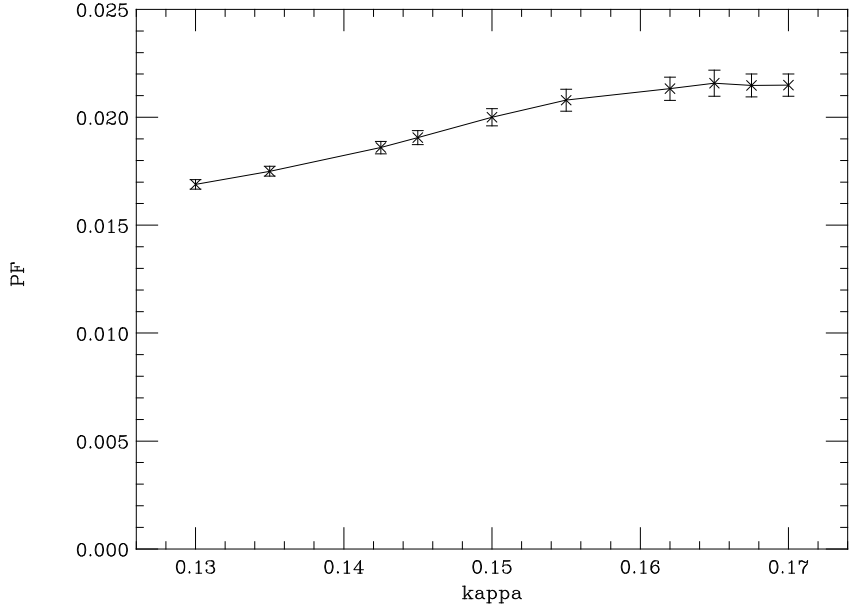}
\caption{Variation of PF with $\kappa$ at $\beta=5.5$ and $h=0.001$.}
\label{fig5}
\end{figure}

 We show in Fig.~\ref{fig5} the variation of the computed order
parameter with $\kappa$ over the range used for $h=0.001$. No sharp
change is indicated as the value of $\kappa_c = 0.16$ is crossed.

Analysis similar to that described above is also performed for this data set.

\begin{figure}
\vspace{3in}
\includegraphics{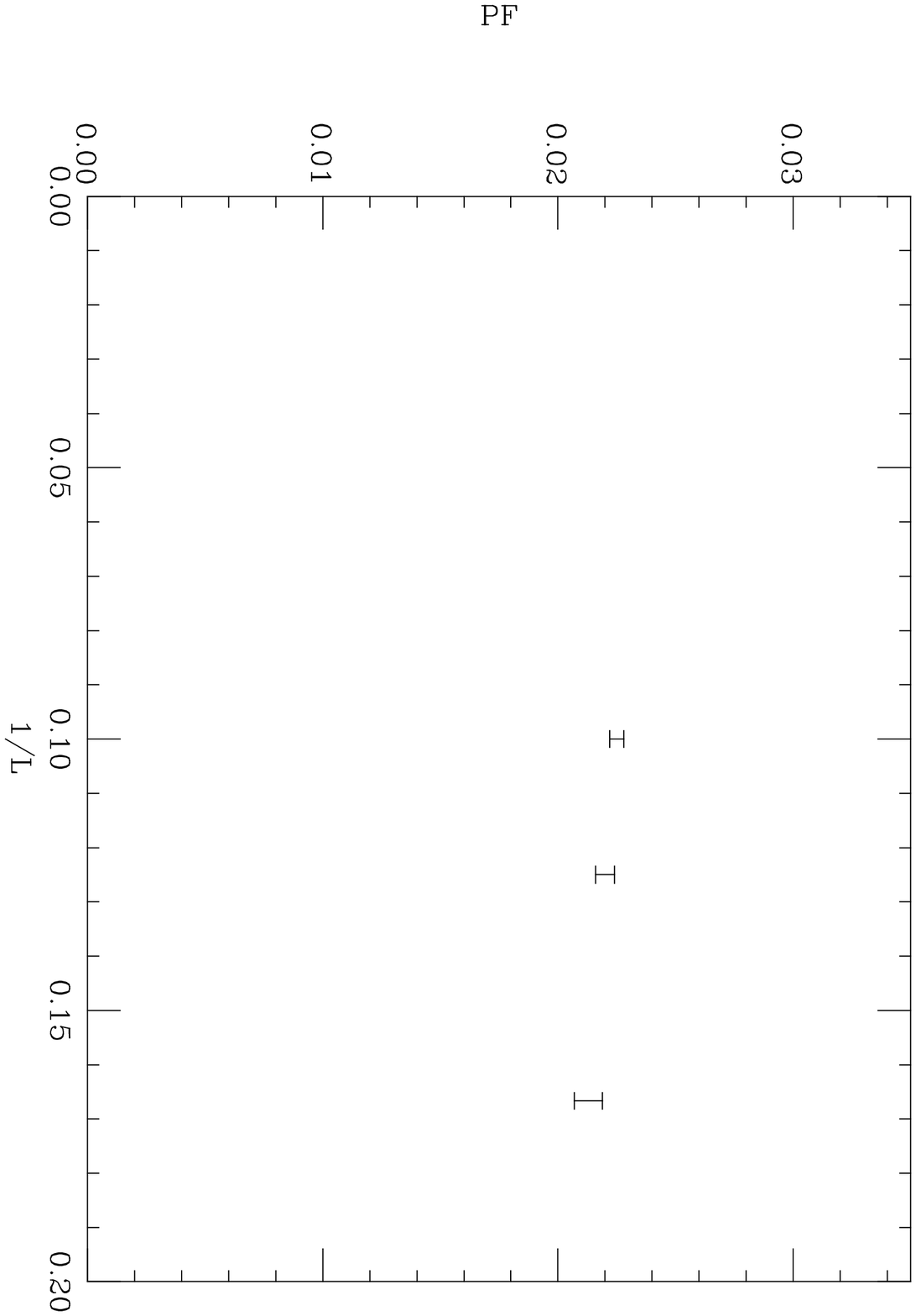}
\includegraphics{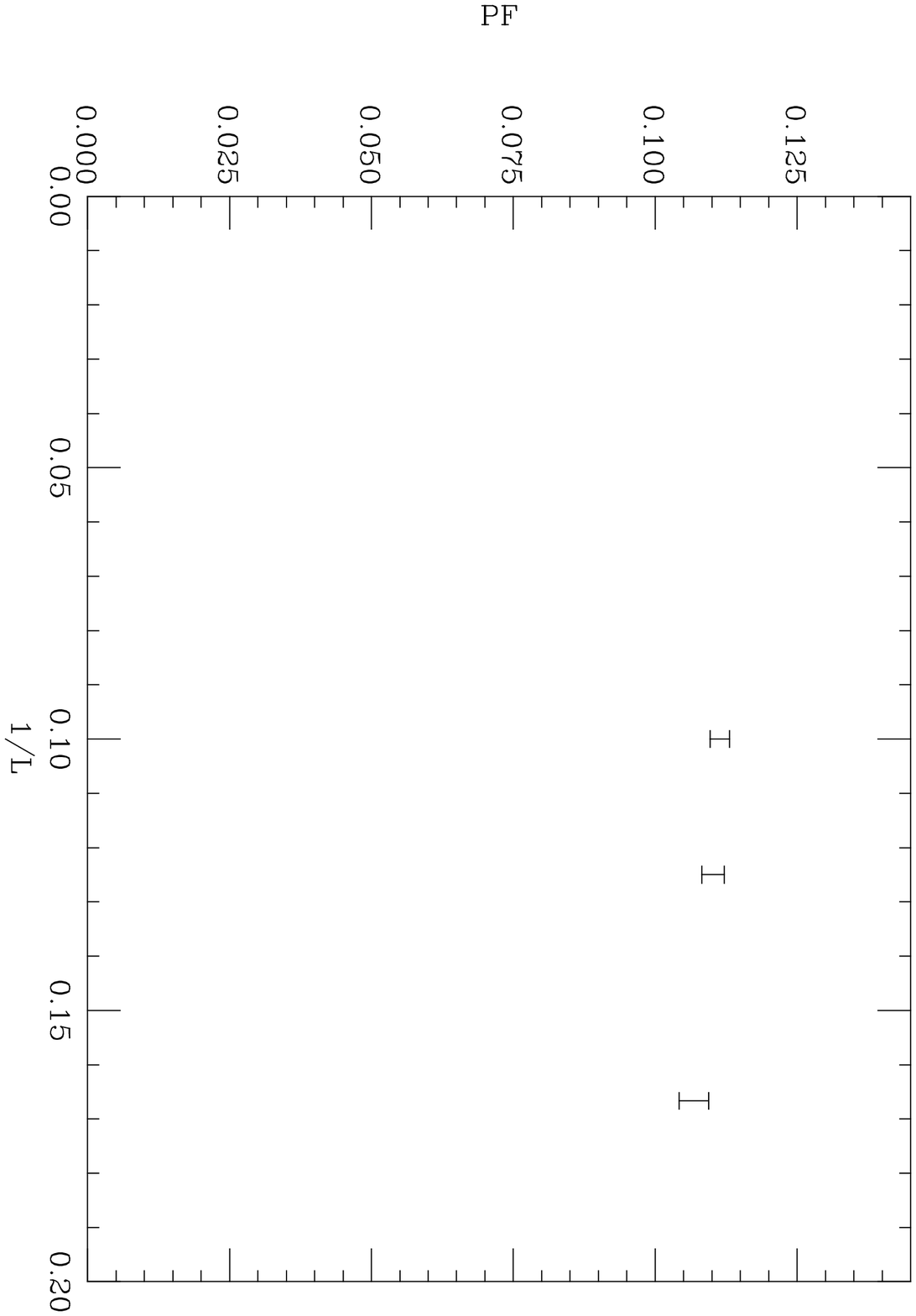}
\caption{(a) PF vs. $1\over L$ for $\beta=5.5$ $\kappa=0.162$ $h=0.001$; and
(b) PF vs. $1\over L$ for $\beta=5.5$ $\kappa=0.162$ $h=0.005$.}
\label{fig6}
\end{figure}

 The results at the larger volumes show only an incremental increase,
if any, as shown, for example, for the case of $\kappa=0.162$ at both
$h=0.001$ and $h=0.005$, in Figs.~\ref{fig6}a and b.

\begin{figure}
\vspace{3in}
\includegraphics{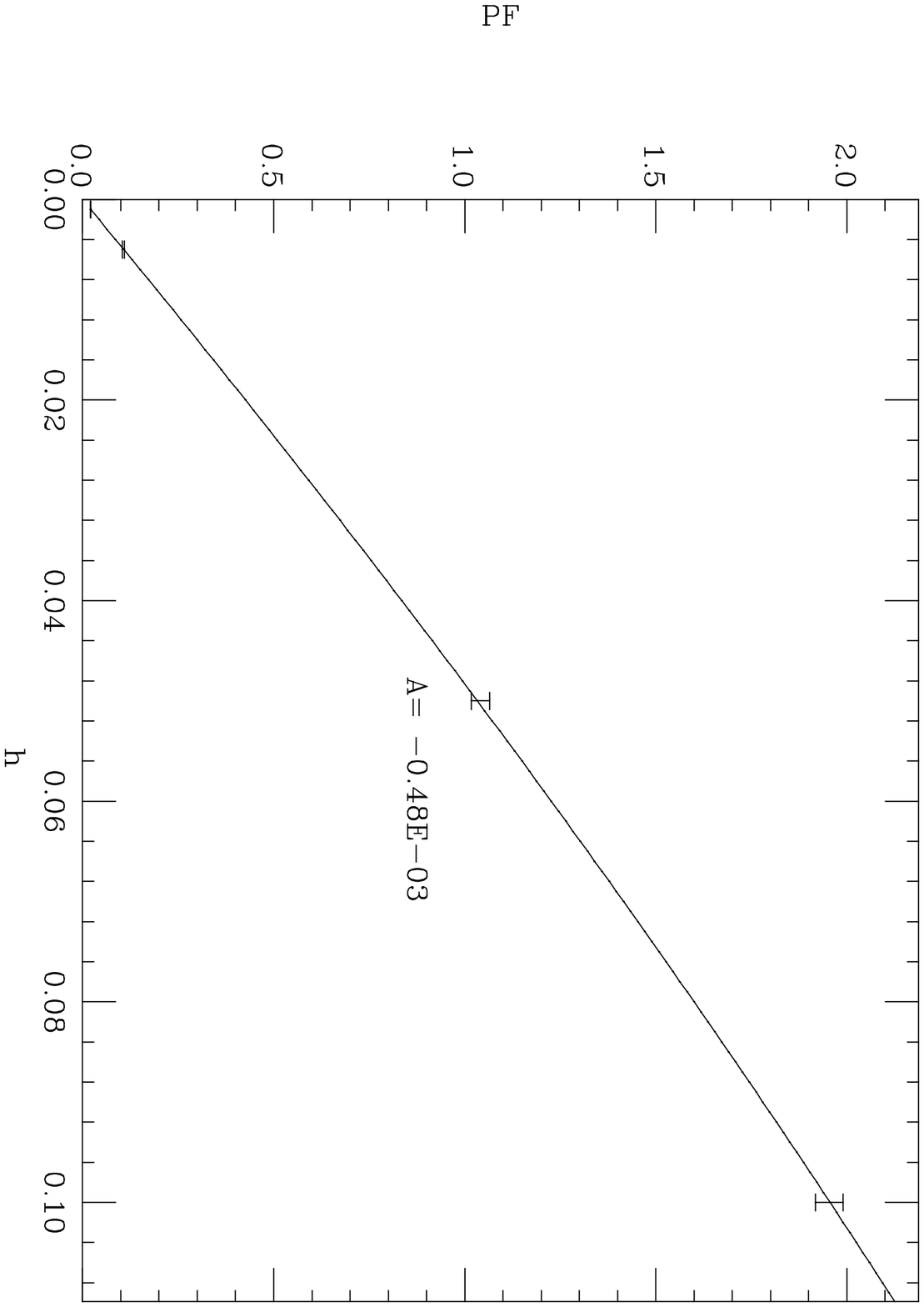}
\includegraphics{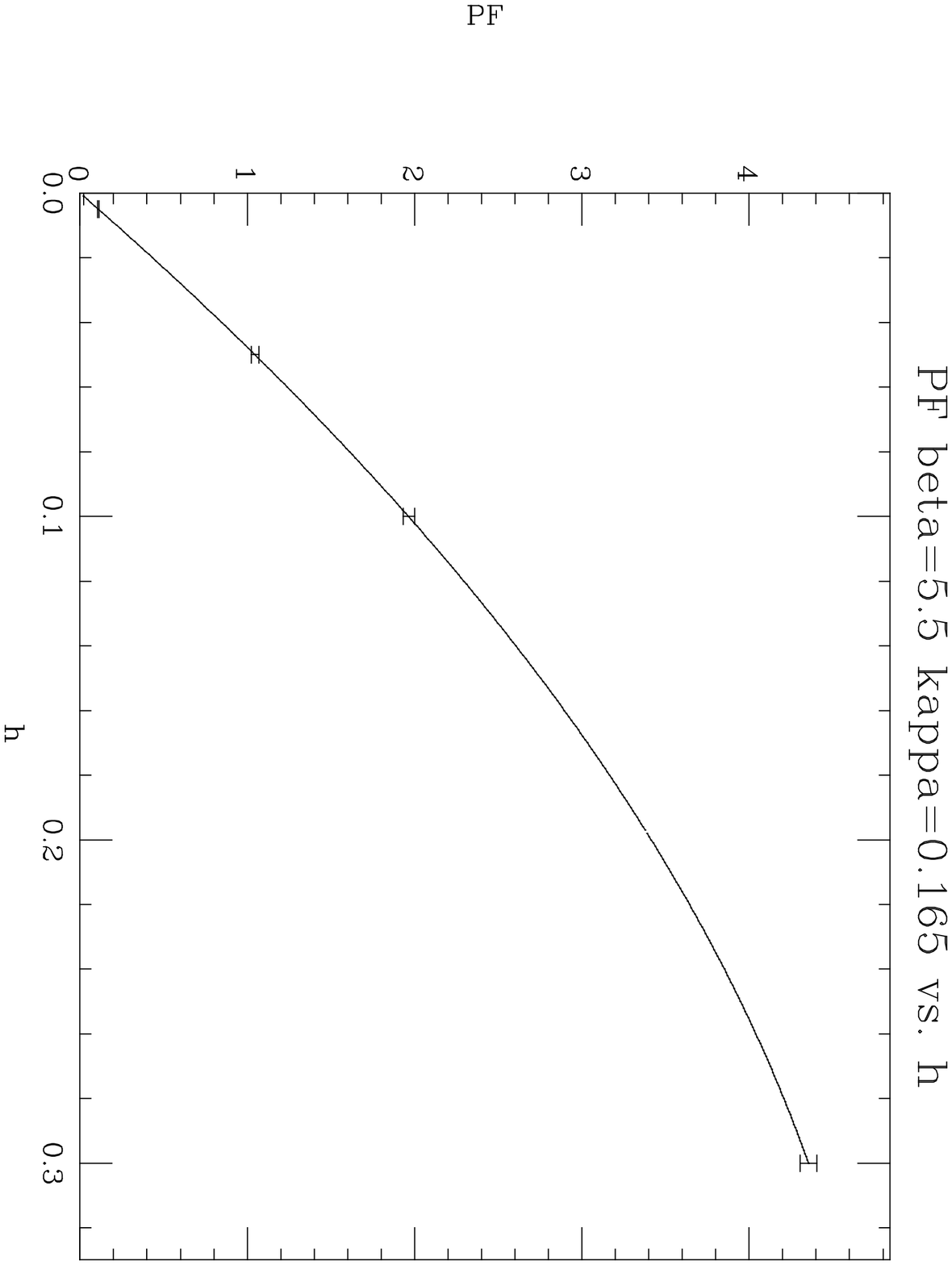}
\caption{(a) PF vs. $h$ for $\beta=5.5$ $\kappa=0.162$ and a quadratic
fit; and (b) PF vs. $h$ for $\beta=5.5$ $\kappa=0.165$ and a quadratic
fit.}
\label{fig7}
\end{figure}

Here again an `infinite volume' limit may indeed be inferred and a
quadratic fit in $h$ gives for the `constant' in the fit a value which
is consistent with zero as shown in Figs.~\ref{fig7}a, and b for
$\kappa=0.162$ and $\kappa=0.165$.

We are then again led to conclude the absence of a parity-flavor 
breaking phase at these values of $\kappa$ above $\kappa_c$.

Attempts at fits with a leading $ h^{1\over 3} $ behaviour again lead
invariably to worse fits indicating again a dominant quadratic
behaviour of  the effective potential for the order parameter. 

\subsection{$\beta=8.0$} 

  The value of $\kappa_c$ in this case has not been determined
numerically. We estimate its value using a tadpole improved
perturbative procedure as discussed in~\cite{Lepage93}. We obtain in
this case a value in the neighborhood of $\kappa_c \simeq 0.145 $.
Consequently our simulations are performed at values of $\kappa$ below
and above this value as shown in Table~\ref{table-three}.

\begin{table}
  \caption{Parameters and measured order parameter $PF_L$ for the
 case of $\beta=8.0$ on Lattices of Volume $ L^4$,  with $L=6, 8, 10$.}
\begin{center}
    \begin{tabular}{ccccc}
      \hline
       $\kappa$ & $h$ & $PF_6$ & $ PF_8$ & $PF_{10}$\\
      \hline
       0.1200 & 0.001 & 0.01591(19) &             &             \\
       0.1200 & 0.005 & 0.0796(10)  &             &             \\
       0.1200 & 0.050 & 0.7845(89)  &             &             \\
       0.1200 & 0.100 & 1.541(17)   &             &             \\
      \hline
       0.1300 & 0.001 & 0.01691(25) & 0.01708(15) & 0.01702(10) \\
       0.1300 & 0.005 & 0.0843(12)  & 0.0851(7)   & 0.0851(7)   \\
       0.1300 & 0.050 & 0.832(10)   &             &             \\
       0.1300 & 0.100 & 1.627(19)   &             &             \\
      \hline
       0.1400 & 0.001 & 0.01782(29) &             &             \\
       0.1400 & 0.005 & 0.0897(18)  &             &             \\
       0.1400 & 0.050 & 0.875(14)   &             &             \\
       0.1400 & 0.100 & 1.732(22)   &             &             \\
      \hline
       0.1460 & 0.001 & 0.01806(30) & 0.01861(20) & 0.01853(14) \\
       0.1460 & 0.005 & 0.0908(16)  & 0.0926(12)  & 0.0927(7)   \\
       0.1460 & 0.050 & 0.892(14)   &             &             \\
       0.1460 & 0.100 & 1.721(25)   &             &             \\
      \hline
       0.1500 & 0.001 & 0.01823(29) & 0.01857(21) & 0.01855(10) \\
       0.1500 & 0.005 & 0.0920(17)  & 0.0928(9)   & 0.0927(6)   \\
       0.1500 & 0.050 & 0.902(14)   &             &             \\
       0.1500 & 0.100 & 1.736(26)   &             &             \\
      \hline
       0.1550 & 0.001 & 0.01838(29) &             &             \\
       0.1550 & 0.005 & 0.0919(15)  &             &             \\
      \hline
       0.1600 & 0.001 & 0.01829(28) &             &             \\
       0.1600 & 0.005 & 0.0921(15)  &             &             \\
       0.1600 & 0.050 & 0.905(13)   &             &             \\
       0.1600 & 0.100 & 1.749(25)   &             &             \\
      \hline
       0.1800 & 0.001 & 0.02012(76) &             &             \\
       0.1800 & 0.005 & 0.0896(12)  &             &             \\
       0.1800 & 0.050 & 0.878(13)   &             &             \\
       0.1800 & 0.100 & 1.703(22)   &             &             \\
      \hline
    \end{tabular}
    \label{table-three}
\end{center}
\end{table}

\begin{figure}
\vspace{3.25in}
\includegraphics{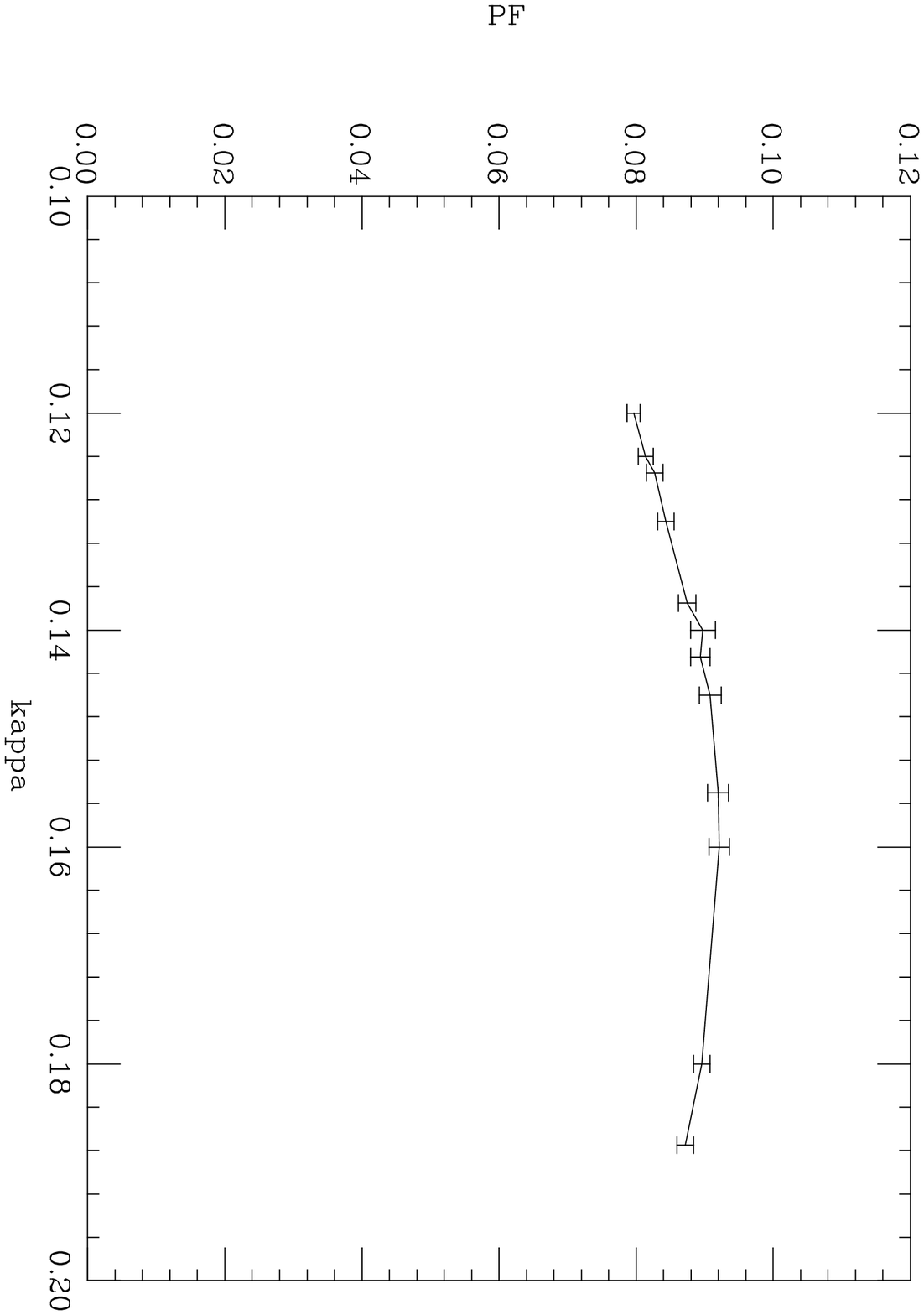}
\caption{Variation of PF with $\kappa$ at $\beta=8.0$ and $h=0.005$.}
\label{fig8}
\end{figure}

 We show in Fig.~\ref{fig8} the variation of the computed order
parameter with $\kappa$ over the range used. No sharp change is
indicated as the value of $\kappa_c \simeq 0.145$ is crossed.

\begin{figure}
\vspace{3in}
\includegraphics{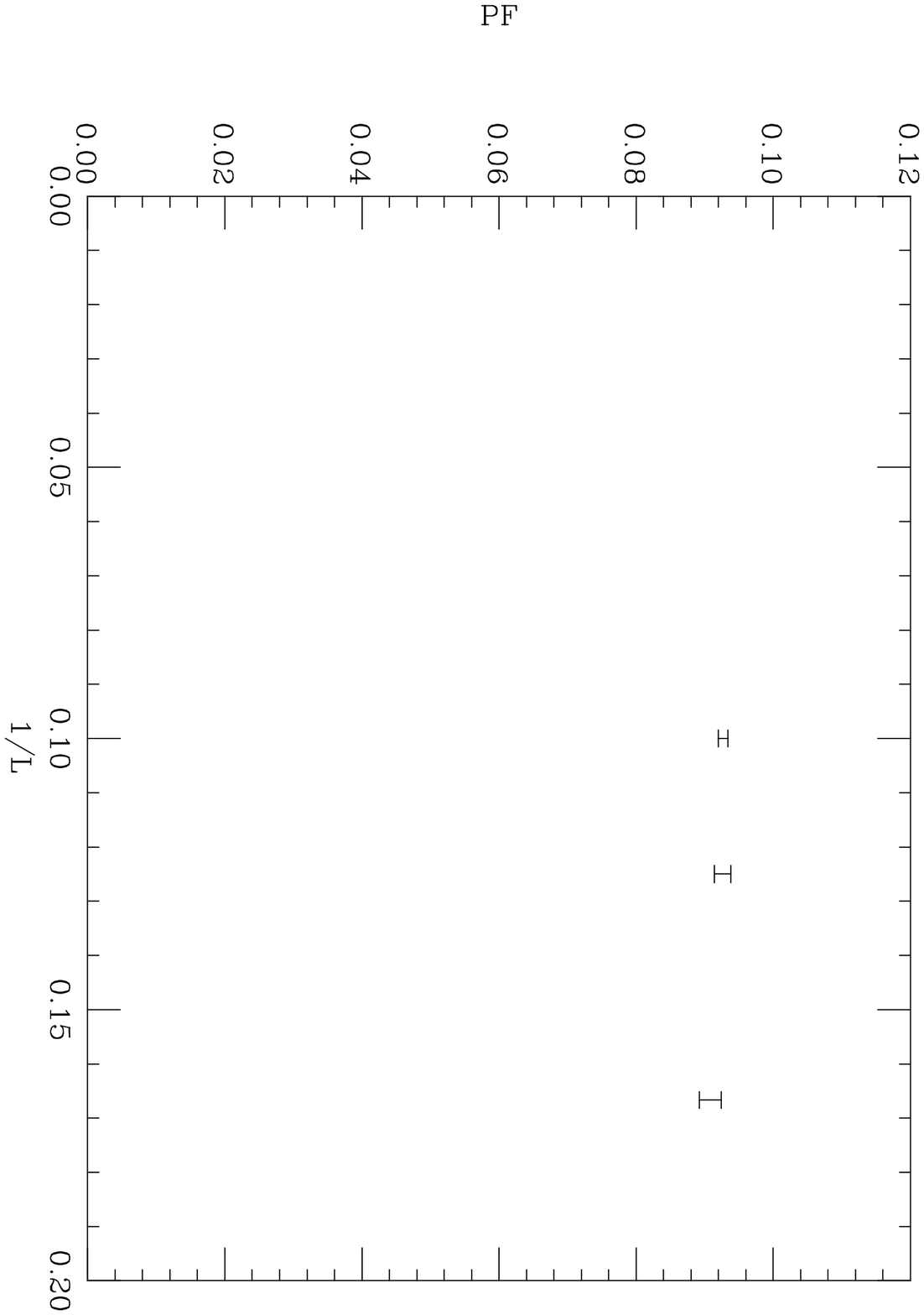}
\includegraphics{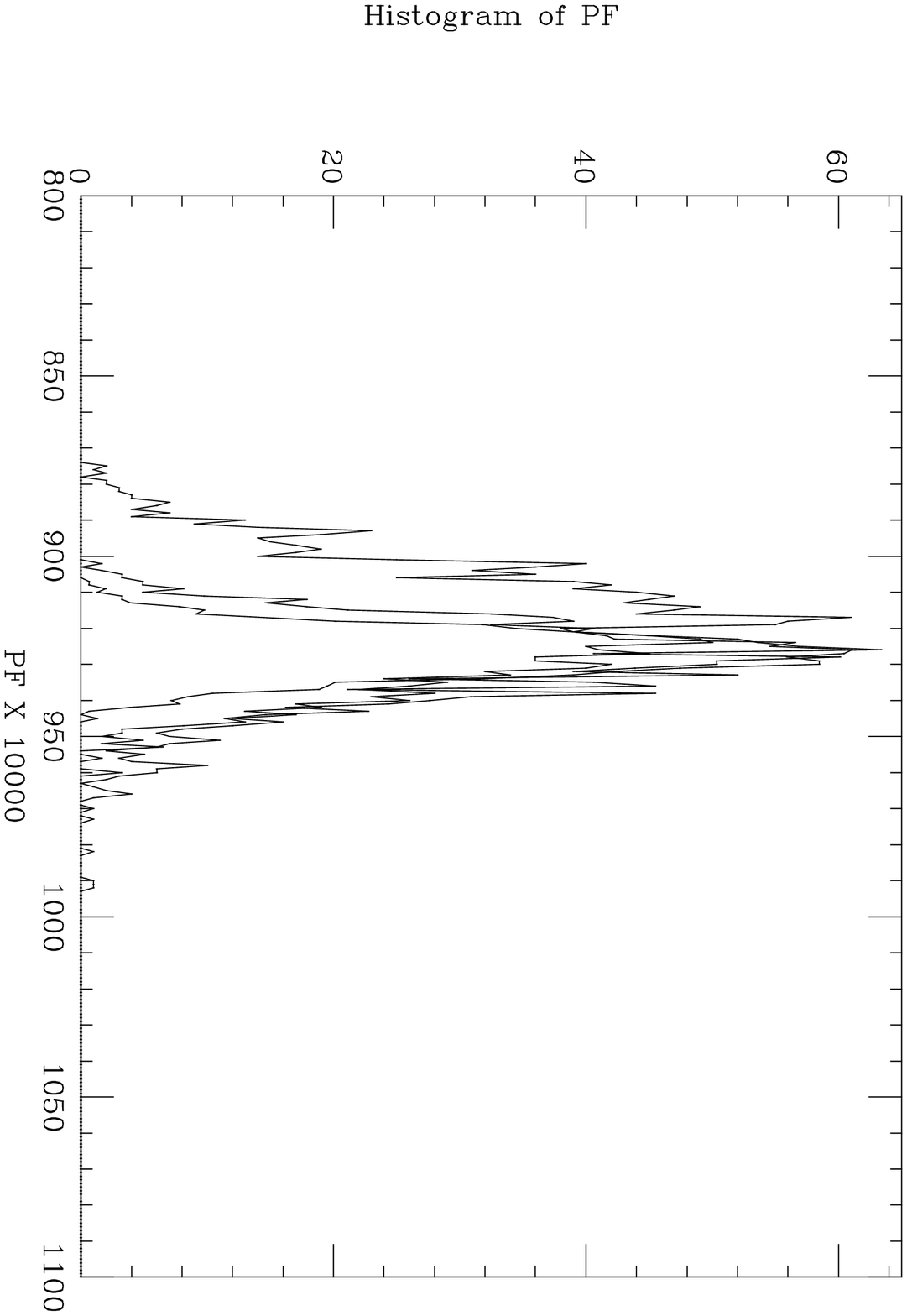}
\caption{(a) PF vs. $1\over L$ for $\beta=8.0$ $\kappa=0.146$
$h=0.005$; and (b) Histogram of computed PF at $\beta= 8.0$ 
$\kappa=0.15$ $h=0.005$ for all volumes considered.}
\label{fig9}
\end{figure}

 We concentrate here on the data at the values of $\kappa $
 above $\kappa_c$. Using the same procedure as above essentially the
same conclusion follows.
 Figs.~\ref{fig9}a and b show that no significant change in the
evaluation of the order parameter at the larger volumes exists for
$\kappa=0.146$ at $h=0.005$ and as seen by the overlapping histograms
for $\kappa=0.15$ at $h=0.005$.

\begin{figure}
\vspace{3in}
\includegraphics{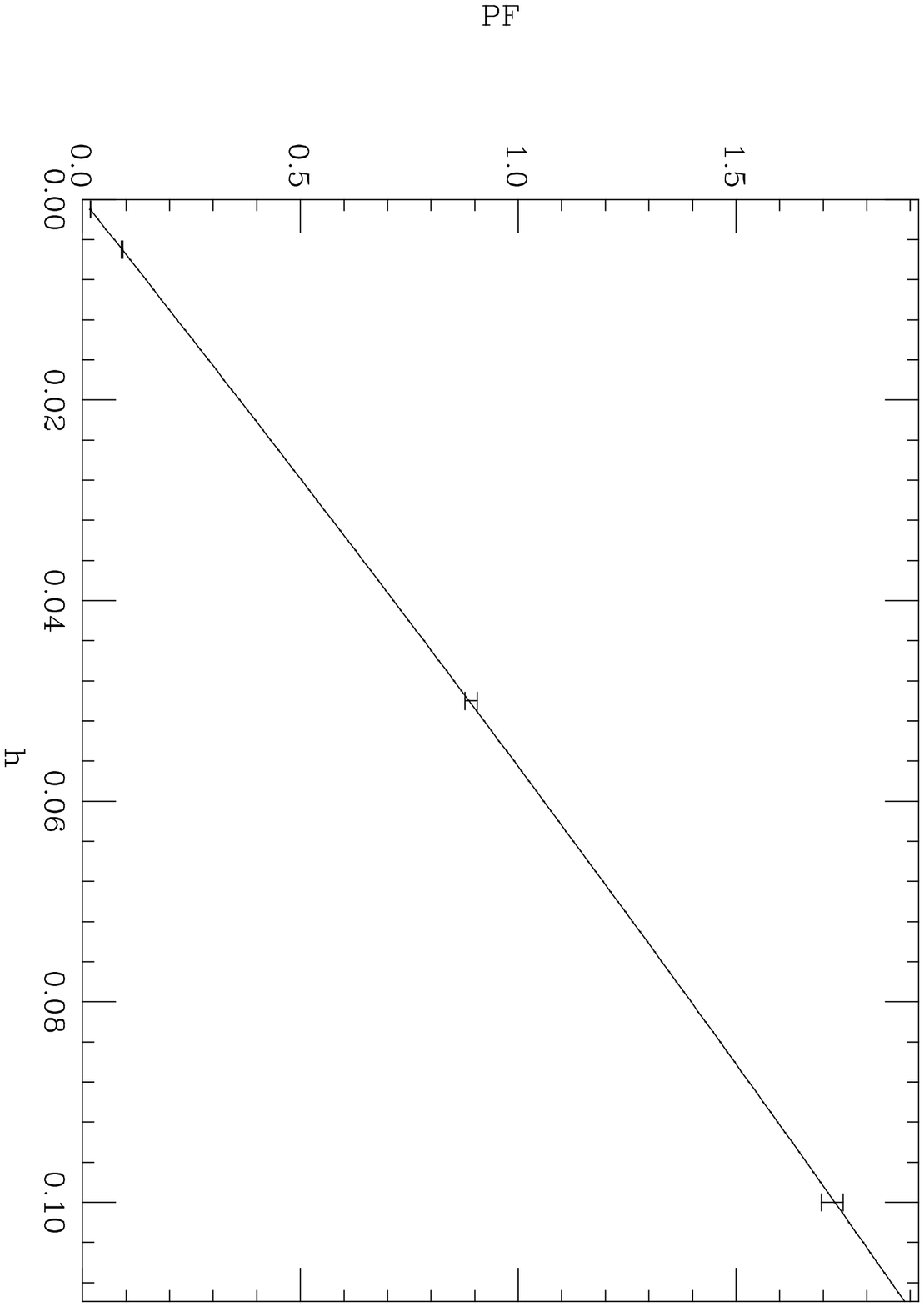}
\includegraphics{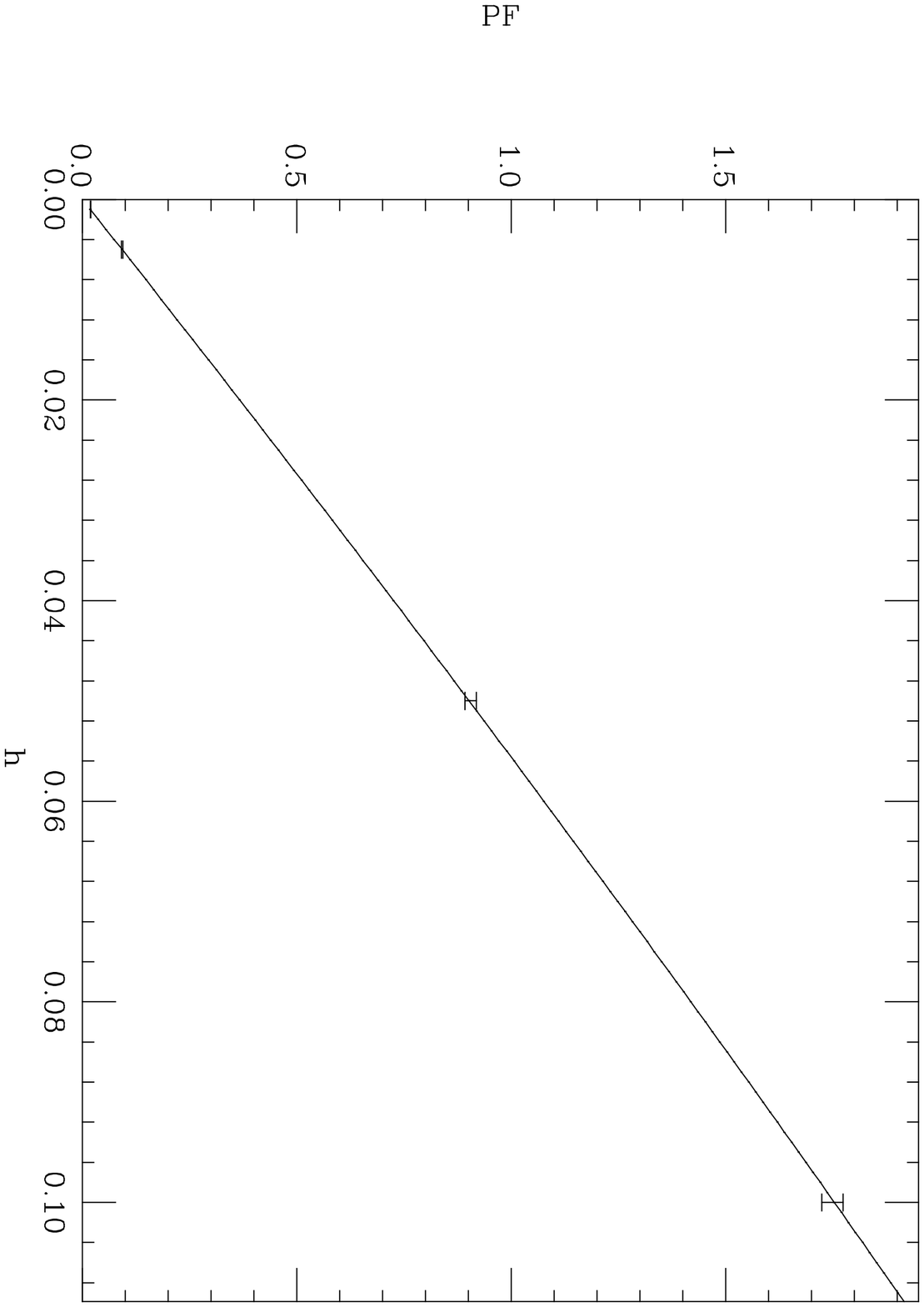}
\caption{(a) PF vs. $h$ for $\beta=8.0$ $\kappa=0.146$ and a quadratic
fit; and (b) PF vs. $h$ for $\beta=8.0$ $\kappa=0.16$ and a quadratic
fit.}
\label{fig10}
\end{figure}

 Furthermore, quadratic fits in $h$ at, for example, $\kappa=0.146 $
and $ \kappa = 0.16$ again have a leading `constant' that is
consistent with zero as shown in Figs.~\ref{fig10}a and b,
respectively.
 Therefore, one is led again to the absence of any signal for a
parity-flavor breaking phase at this value of $\beta$.
 
Finally the obvious leading linear dependence of the fit indicates
again a dominant quadratic effective potential at this value of
$\beta$ as well.

\section{Conclusions}
\label{Conclusions}

It is clear from the discussion above that QCD with two flavors of
Wilson fermions does not exhibit a parity-flavor breaking phase at
$\beta >5.0 $ as postulated by Aoki and collaborators. Since it has
been demonstrated that at $\beta=0.0$ such a phase may exist in a
large $N$ (color) limit, it is also clear that if this phase does
extend beyound $\beta=0.0$, it must pinch out in a manner similar to
that in the NJL model at  $\beta < 5.0$. In either case this phase
would not be relevant for the discussion of the approach to the chiral
limit in QCD and the ensuing Goldstone nature of the pions for $\beta
> 5.0$. In fact, all indications are such that, as shown formally
sometime ago, this is simply related to the approach to zero lattice
spacing and infinite volume.

\section*{Acknowledgements}
I wish to thank Urs Heller for providing a modified QCD code for
performing the simulations reported in this paper. All these
simulations were done on the SCRI IBM compute cluster. This research
was supported by the U.S. Department of Energy through Contract
Nos. DE-FG05-92ER40742 and DE-FC05-85ER250000.

\bibliographystyle{unsrt}

\begin{thebibliography}{9}
\bibitem{Aoki84a}
 S. Aoki,  Phys. Rev. D30, 2653 (1984)
\bibitem{Aoki84b}
 S. Aoki,  Phys. Rev. D33, 2339 (1984)
\bibitem{Aoki86}
 S. Aoki,  Phys. Rev. Letters 57, 3136 (1986)
\bibitem{Aoki89}
 S. Aoki,  Nucl. Phys. B314, 79,  (1989)
\bibitem{Aoki95}
 S. Aoki,  On the phase structure of QCD with Wilson Fermions, 
 UTHEP-318,  August (1995)
\bibitem{Gocksch89}
 S. Aoki  and A. Gocksch. Phys. Letters B231, 449 (1989)
\bibitem{Gocksch90}
 S. Aoki  and A. Gocksch. Phys. Letters B243, 409 (1990)
\bibitem{Gocksch92}
 S. Aoki  and A. Gocksch. Phys. Rev. D45,  3845 (1992)
\bibitem{Vafa84a}
 C. Vafa and E. Witten, Nuclear Physics B234, 173 (1984)
\bibitem{Vafa84b}
 C. Vafa and E. Witten, Phys. Rev. Letters 53, 535 (1984)
\bibitem{Karsten81}
 L.H. Karsten and J. Smit,  Nucl. Phys. B183,  103 (1981)
\bibitem{Seiler82}
 E. Seiler and I. Stamatescu,  Phys. Rev. D25,  2177 (1982)
\bibitem{Bichicchio85}
 M. Bochicchio,  L. Maiani,  G. Martinelli,  G. Rossi and M. Testa, 
  Nuclear Physics B262,  331 (1985)
\bibitem{Bitar94a}
 K. M. Bitar and P. Vranas, Phys. Rev D50,  3406 (1994)
\bibitem{Bitar94b}
 K. M. Bitar and P. Vranas, Nuclear Physic B,  Proc. Suppl. 34, 661 (1994)
\bibitem{Gocksch94}
 S. Aoki,  S. Boettcher and A. Gocksch,  Phys. Letters B331, 157 (1994)
\bibitem{Horvath95}
 I. Horvath, The Phase Structure of the Schwinger Model on the Lattice
with Wilson Fermions in the Hartree-Fock Approximation, FSU-SCRI-95-101;
to appear in Phys. Rev. D (1996), 
\bibitem{Creutz95}
 M. Creutz,  BNL-62123,  May 1995
\bibitem{Lepage93} 
P. Lepage and P.G. Mackenzie,  Phys. Rev. D48, 2250 (1993)

\end{thebibliography}

\end{document}